\documentclass[a4paper,fleqn]{cas-sc}

\usepackage[numbers]{natbib}
\usepackage{textgreek}
\usepackage{commath}

\def\tsc#1{\csdef{#1}{\textsc{\lowercase{#1}}\xspace}}
\tsc{WGM}
\tsc{QE}

\begin{document}
\let\WriteBookmarks\relax
\def\floatpagepagefraction{1}
\def\textpagefraction{.001}

\shorttitle{The effects of fair allocation principles on energy system model designs}    

\shortauthors{Vågerö, Zeyringer and Inderberg}  

\title [mode = title]{The effects of fair allocation principles on energy system model designs}  

\author[1]{Oskar Vågerö}[orcid=0000-0002-0806-0329]

\cormark[1]
\ead{oskar.vagero@its.uio.no}
\credit{
    Conceptualisation, 
    Data curation,
    Formal analysis,
    Investigation
    Methodology, 
    Validation, 
    Visualisation,
    Writing - original draft
}

\author[2]{Tor Håkon Jackson Inderberg}[orcid=0000-0002-1838-3834]
\ead{thinderberg@fni.no}
\credit{
    Conceptualization,
    Supervision,
    Writing - original draft
}

\author[1]{Marianne Zeyringer}[]
\ead{marianne.zeyringer@its.uio.no}
\credit{
    Conceptualisation,
    Methodology,
    Supervision,
    Writing - original draft
}

\affiliation[1]{organization={Department of Technology Systems, University of Oslo},
            addressline={Gunnar Randers vei 19}, 
            city={Kjeller},
            postcode={N-2007}, 
            country={Norway}}

\affiliation[2]{organization={Fridtjof Nansen Institute},
            addressline={Fridtjof Nansens vei 17}, 
            city={Lysaker},
            postcode={N-1326}, 
            country={Norway}}

\cortext[1]{Corresponding author}

\begin{abstract}
    What constitutes socially just or unjust energy systems or transitions can be derived from the philosophy and theories of justice. Assessments of distributive justice and utilising them in modelling lead to great differences based on which justice principles are applied. From the limited research so far published in the intersection between energy systems modelling and justice, we find that comparisons between the two principles of utilitarianism and egalitarianism dominate in assessments of distributive justice, with the latter most often considered representing a 'just energy system'. The lack of recognition of alternative and equally valid principles of justice, resting on e.g. capabilities, responsibilities and/or opportunities, leads to a narrow understanding of justice that fails to align with the views of different individuals, stakeholders and societies. More importantly, it can lead to the unjust design of future energy systems and energy systems analysis. 
    
    In this work, we contribute to the growing amount of research on distributive justice in energy systems modelling by assessing the implications of different philosophical views on justice on modelling results. Through a modelling exercise with a power system model for Europe (highRES), we explore different designs of a future (2050) net-zero European electricity system, and its distributional implications based on the application of different justice principles. In addition to the utilitarian and egalitarian approach, we include, among others, principles of 'polluters pay' and 'ability-to-pay', which take historical contributions of greenhouse gas emissions and the socio-economic conditions of a region into account. 
    
    We find that fair distributions of electricity generating infrastructure look significantly different depending on the justice principles applied. The results may stimulate a greater discussion among researchers and policymakers on the implications of different constructions of justice in modelling, expansion of approaches, and demonstrate the importance of transparency and assumptions when communicating such results.
\end{abstract}



\maketitle

\section{Introduction}\label{sec:introduction}
Limiting global temperature increase and climate change requires rapid and deep decarbonisation of our economies and energy systems. Such profound changes are not merely techno-economic, but also include socio-cultural changes and challenges \cite{perlaviciutePerspectiveHumanDimensions2021}. In addition to efforts of transitioning to low-carbon energy systems, the importance of just transitions which equitably distributes benefits and addresses potential injustices has received increased attention both in academic literature \cite{mccauleyJustTransitionIntegrating2018,carleyJusticeEquityImplications2020}, and in policy discussions \cite{europeancommissionEuropeanGreenDeal2019,rep.ocasio-cortezTextRes1092019}. The concept of just transitions acknowledges that the shift towards low-carbon energy systems may create new inequalities or amplify existing ones, affecting different communities in diverse ways. When studying energy transitions, there is a need to include a wide range of justice aspects.

Energy systems optimisation modelling (ESOM) provides a tool to analyse energy system transitions and future energy systems. It is often used to advise policymakers and planners on how to best design energy systems in line with policy goals such as the European Union Green Deal, which aims to transform the European Union to be climate neutral and just by 2050. While ESOMs are strong tools to support the first objective focusing on techno-economic aspects, they  often overlook socio-cultural aspects \cite{krummModellingSocialAspects2022,sovacoolIntegratingSocialScience2015} required to analyse just energy systems. However, the state-of-the-art is developing, and more attention is devoted to the inclusion of various socio-cultural aspects \cite{krummModellingSocialAspects2022,hirtReviewLinkingModels2020,pfenningerEnergySystemsModeling2014} such as socio-political acceptance of  energy infrastructure \cite{bolwigClimatefriendlySociallyRejected2020,flachsbarthAddressingEffectSocial2021,grimsrudSpatialTradeOffsNational2023,koecklinPublicAcceptanceRenewable2021,weinandImpactPublicAcceptance2021,priceImplicationsLandscapeVisual2022,tsaniOutSightOut2024}, behaviour and lifestyle changes \cite{lawrenzExploringEnergyPathways2018,lombardiMultilayerEnergyModelling2019,bartholdsenPathwaysGermanyLowCarbon2019,burandtDecarbonizingChinaEnergy2019,auerDevelopmentModellingDifferent2020,pergerPVSharingLocal2021}, and spatial distribution of benefits and burdens \cite{drechslerEfficientEquitableSpatial2017, sasseDistributionalTradeoffsRegionally2019, sasseRegionalImpactsElectricity2020,fellCapturingDistributionalImpacts2020,neumannCostsRegionalEquity2021,chenBalancingGHGMitigation2022,pedersenUsingModelingAll2023,sasseLowcarbonElectricitySector2023}. These examples only represent a small share of the studies applying ESOMs and there is considerable room to further advance the field. 

Optimisation modellers have as such sought new tools and techniques to attend to non-technical aspects of energy systems, often driven by the wish to include political feasibility and fairness of the system. Perceptions of justice and fairness are found to be central to the well functioning of society and perceived unfair outcomes can result in protests, damaged relationships and divided communities \cite{grossCommunityPerspectivesWind2007}. Most previous research have focused on spatial distribution of energy infrastructure \cite{vageroCanWeOptimise2023}, often as a proxy for socio-political feasibility. One of these techniques, Modelling to Generate Alternatives (MGA), allows modellers to investigate the structural uncertainty and solutions which are near-optimal through an iterative modelling-process \cite{decarolisUsingModelingGenerate2011}. Previous studies applying MGA have explored alternative energy system designs that are slightly more costly, but tend to other objectives. For example, Sasse \& Trutnevyte looked at the trade-offs between cost-efficiency and a just distribution of energy infrastructure in Switzerland \cite{sasseDistributionalTradeoffsRegionally2019}, Lombardi et al. mapped out spatially and technologically diverse system configurations for Italy \cite{lombardiPolicyDecisionSupport2020} and Chen, Kirkerud \& Bolkesjø analysed alternative designs for a Northern European energy system with attention to reduced land-use conflicts \cite{chenBalancingGHGMitigation2022}. 

Throughout these and other examples of optimisation modelling studies, there is generally a narrow interpretation and limited discussion of what entails a just distribution, which in turn may result in equally narrow policy recommendations \cite{vageroCanWeOptimise2023}. In fact, this issue persists beyond modelling and Van Uffelen et al. \cite{vanuffelenRevisitingEnergyJustice2024} have found that research applying the energy justice framework most often does not explicitly support normative claims or recommendations and that implicit principles of justice are a source of normative uncertainty.

Furthermore, utilitarianism and neoclassical economics exclusively dominate the field of ESOM, leading to depictions of ‘non-normative’ (cost-optimal) versus approaches that account for different aspects of justice or fair distribution. This, however, disregards the fact that also the conventional cost-optimal approach comes with significant justice implications. Following the frequently quoted expanded aphorism “all models are wrong, but some are useful” \cite{boxScienceStatistics1976}, modelling is often concerned with non-epistemic values such as usefulness \cite{silvastWhatEnergyModellers2020}. As such cost-optimal modelling is indeed normative and basing, for example, policy recommendations on it implicitly include the (normative) assumption that inequities within the system are unimportant (or at least that efficient redistribution will adjust the inequities). We will later refer to this as 'efficiency-based justice'. Sgouridis et al. extends the saying to “all models are subjective, but few acknowledge it” \cite{sgouridisVisionsModelsEthos2022}, highlighting that implicit assumptions create blind spots in models impacting the results and in extension potentially also energy policy.

Other than Pedersen et al., who expand the number of justice interpretations when studying effort-sharing and the distribution of CO\textsubscript{2} abatement cost in a European power system \cite{pedersenUsingModelingAll2023}, modelling analyses are generally limited to compare a cost-optimal baseline to some variant of equal distribution based on population or consumption patterns \cite{drechslerEfficientEquitableSpatial2017,sasseDistributionalTradeoffsRegionally2019,neumannCostsRegionalEquity2021}. The aim of this work is to showcase how the application of different justice principles radically influences electricity system designs, and thereby emphasise the need to explicitly state assumptions and subjectivity when presenting modelling results. To study this, we use MGA as a technique to sample the near-optimal decision space for the design of a European electricity system in 2050, which we then analyse quantitatively based on differently defined Gini coefficients as a proxy for the distributional justice of a system configuration. To better understand the implications for future electricity systems, we also compare this with the current and cost-optimal electricity system. While our work is confined to the field of electricity systems modelling, the analysis is likely to be applicable also for other settings where justice or distribution needs to be defined.  
 
The remainder of this paper is structured as follows: Section \ref{sec:theory} presents the conceptual framework of the work, focusing on different principles of justice and methods of allocating benefits and burdens. Section \ref{sec:method} outlines the methodology, including a brief description of the linear optimisation electricity system model highRES, the quantification of justice through the Gini coefficient and the MGA technique applied. Section \ref{sec:results} contains the results of the modelling exercise, which are further discussed in section \ref{sec:discussion}. 

\section{Principles and methods of just allocation}
\label{sec:theory}
We apply seven diverging justice principles for just distribution of electricity infrastructure. The criteria for selection was that they appear in the scientific literature, and that the data required is available and suitable for a European electricity system model. See table \ref{tbl:principles} for exemplary literature. These seven approaches to justice are necessarily quite stylised in their presentational form. This is because of the need for brevity given the purpose of being applicable in the modelling, and while we have provided some links to their origin and links to the literature, we have emphasised concrete presentations. Nor is the list exhaustive, and alternatives could have been included, based on, for example, bio- or ecocentrism. However, these seven approaches include key interpretations which can be found in scientific literature and at the heart of political philosophy (see e.g. Kverndokk and Rose \cite{kverndokkEquityJusticeGlobal2008}, Zhou and Wang \cite{zhouCarbonDioxideEmissions2016} or Zimm et al. \cite{zimmJusticeConsiderationsClimate2024}).

\subsection{Efficiency-based justice}
Efficiency-based distribution is commonly the core principle in energy modelling. Also known as utilitarianism, it has a basis in ethical theory, often traced back to Jeremy Bentham \cite{benthamIntroductionPrinciplesMorals1789} and John Stuart Mill \cite{millUtilitarianism1863}. In its simplest definition, utilitarianism claims that we should seek to maximise human welfare, or utility, for members of society \cite{kymlickaContemporaryPoliticalPhilosophy2002}. Traditionally, utility has been defined in terms of happiness, pleasure, or the absence of pain, but the interpretation can be context-specific. In the context of electricity systems planning, utility can be understood as access to energy-related services, which in turn relies on efficient resource allocation to ensure supply meets or exceeds demand, ultimately leading to the largest monetary surplus for society as a whole. However, the pursuit of maximising overall utility, sometimes through pareto optimisation, often leads to unequal distributions of utility among different members of society, leading to perceptions of unjust distributions at individual, group, or geographic levels. 

The debate over whether utilitarianism justifies treating people as mere means to an end, such as sacrificing the weak for the sake of the majority, remains a frequent criticism. While utilitarianism was originally viewed as progressive and reform-oriented, it is now often considered preserving the status quo \cite{kymlickaContemporaryPoliticalPhilosophy2002}. 

\subsection{Equality-based justice}
Equality, or egalitarianism, on the contrary, is based on the fundamental principle that all people should have equal rights, opportunities and treatment, irrespective of their background, social status or other characteristics. This includes not only equal access to resources but also the fair and just distribution of benefits and burdens among individuals. A variant of this was posited by John Rawls, inequalities should only be accepted if cases where the worst-off groups will be better off than they were under an equal distribution \cite{rawlsPrinciplesJustice1971,rawlsJusticeFairnessRestatement2001}. With equal treatment of individuals, the distribution of benefits and burdens scale with population, meaning that, for example, a more populous country or region should bear a larger share of both the benefits and burdens. 

Moreover, it is important to distinguish between equality of opportunity and equality of outcome within the framework of egalitarianism. Equality of opportunity holds that inequalities may be acceptable if everyone has the same opportunities and any resulting inequality is a consequence of individual choices rather than societal circumstances \cite{kymlickaContemporaryPoliticalPhilosophy2002}. On the other hand, equality of outcome focuses on the resulting distribution itself, irrespective of the choices or processes leading to it, where perceived fairness of how costs and rewards are distributed across groups and group members \cite{forsythConflict2010}. In the context of electricity systems planning, it is natural to focus on equality of outcome, given that we are examining a static system rather than its development over time. Equality-based justice, however, does not take into account contextual factors, such as access to resources or economic capability.

\subsection{Capability-based allocations}
A different aspect on achieving a fair and just distribution of benefits and burdens is recognising that while individuals deserve equal rights, their opportunities and starting points may differ significantly \cite{senEqualityWhat1979,senIssuesMeasurementPoverty1979}. Consequently, it becomes necessary to consider individuals’ capabilities, or ‘ability to pay’ when distributing benefits, burdens and responsibilities, taking into account the varying circumstances people find themselves in. For instance, individuals or communities with lower economic capabilities may require a different allocation of benefits to address their disadvantaged position in society, and similarly have a lower responsibility to act, compared to groups with higher capabilities. 

While there are multiple ways to define capabilities, one commonly used approach at the country-level is based on economic indicators, such as the evaluation of gross domestic product (GDP) or purchasing power parity (PPP)-adjusted GDP \cite{kverndokkEquityJusticeGlobal2008,zhouCarbonDioxideEmissions2016,kverndokkClimatePoliciesDistributional2018,stephensonEnergyCulturesNational2021}. However, it is important to note that economic capability is just one dimension among many that need to be considered in a comprehensive assessment of individuals’ capabilities, especially when going into more detailed and individual assessments beyond country-level averages. A capability-based allocation can, however, be critiqued for only considering existing capabilities, omitting perspectives of historical developments and changes to countries financial situation.

\subsection{Historic responsibility as justice}
Incorporating a temporal dimension is justice seen as historic responsibility. This perspective argues that previous actions and cumulative CO\textsubscript{2} emissions should serve as the basis for distribution \cite{shueHistoricalResponsibilityHarm2015}. It recognises that certain individuals, communities or countries have historically made larger contributions to the accumulation of greenhouse gas emissions, resulting in the current climate crisis.

While there are counterarguments, such as that the consequences of emitting greenhouse gases was not known at the time of emission, this view represents an argument that the distribution of responsibility, benefits and burdens should be adjusted to account for different levels of responsibility for climate change \cite{jonesNationalContributionsClimate2023,friedlingsteinGlobalCarbonBudget2022}. It suggests that those who have significantly contributed to carbon emissions have a greater obligation to bear the burdens and take responsibility for rectifying the environmental damage caused. Consequently, countries with higher historical emissions might be required to make more substantial efforts in terms of emission reductions, mitigation strategies and financial contributions. 

Responsibility can be allocated based on either production-based or consumption-based emissions, where the former is the common basis within the UNFCCC and the Paris Agreement \cite{stephensonEnergyCulturesNational2021,jonesNationalContributionsClimate2023}. Production-based emissions refer to greenhouse gas emissions produced within a specific geographical area, and resulting activities within that defined boundary. Imported products and materials are as such not covered. On the other hand, consumption-based emissions stem from consumption of goods or services associated with greenhouse gas emissions. This approach encompasses the entire lifecycle emissions and assigns them to the geographical area consuming the goods or services. 

Furthermore, the system boundary may cover different parts of the economy, such as only the energy or electricity sector, leading to different outcomes. 

\subsection{The self-sufficiency principle}
The principle of equality, which asserts that individuals should have equal access to resources, can be further nuanced through acknowledging the differences in energy needs for different geographical areas. Areas with more extreme temperatures, whether hot or cold, generally has different energy consumption needs compared to other areas. This can justify a larger claim to energy resources. To account for this, we, can apply the concept of regional self-sufficiency, whereby countries or regions strive to produce as much energy as they consume \cite{mongsawadPhilosophySufficiencyEconomy2010}. This self-sufficiency principle, particularly in electricity systems, has practical implications and can be motivated by factors such as ensuring security of energy supply or pursuing energy autarky \cite{neumannCostsRegionalEquity2021,trondleTradeOffsGeographicScale2020}. Similar to the equality principle, the self-sufficiency principle does not take contextual factors into account, such as access to natural resources or land area available.

\subsection{Land burden as justice principle} 
Distribution based on land area is a method of allocating resources based on the geographic size of a region and has its origin in land-use conflicts, particularly for renewables  \cite{chenBalancingGHGMitigation2022,scheidelEnergyTransitionsGlobal2012,ioannidisReviewLandUse2020}. There are important counter-arguments to this approach. For example, a distribution solely based on land area does not account for factors like population density, natural resource availability, or economic productivity. A country with a vast land area but a small population might not require the same level of resources as a densely populated area with smaller land area. Similarly, regions with abundant natural resources or high agricultural productivity might not need additional resources simply due to their land area. This aligns well with previous research \cite{christWindEnergyScenarios2017}, which has identified that the available area and population density of a region was deemed important in determining the appropriate societal and ecological burden level from wind power deployment. 

Moreover, when considering the burden of developing additional energy infrastructure, along with recognising the equal right to a natural environment for everyone, it becomes apparent that larger territories might bear the potential of sacrificing more land area, given its abundant nature.

\subsection{The grandfathering principle}
Grandfathering refers to a practice that allows certain individuals or entities to be exempted from new rules, regulations or requirements based on their pre-existing status or conditions \cite{damonGrandfatheringEnvironmentalUses2019}. Essentially, it grants certain privileges, rights or exemptions to those who were already engaged in a particular activity or possessed certain qualifications before the new rules or requirements came into effect. According to the principle, past efforts and developments should be continued and valued, resembling a conservative approach. The principle is often considered in real-life climate negotiations, such as that of the Kyoto Protocol and the EU ETS \cite{knightWhatGrandfathering2013}. It suggests that past emissions, to some extent, entitle entities to further emissions. This principle aligns with the aim of maintaining established economic and social activities, and argues that sudden and radical transitions would be socially unjust. In contrast, it can be seen as opposing the concept of historic responsibility. 

Even if grandfathering may be disputed as a justice principle \cite{caneyJusticeDistributionGreenhouse2009} and is sometimes seen as a result of pragmatism and political negotiation \cite{meyerClimateJusticeHistorical2010}, we still see a value in including it in the analysis. Being an oft-applied principle in practice (and in scientific studies \cite{chenRegionalEmissionPathways2021}), grandfathering is important in that it values current economic activities, and recognises that undermining of these is unjust, or can have detrimental effects on citizens in some regions. While it runs the risk of conserving unfavourable economic activities  and can lead to the continuation of existing advantages for certain stakeholders, it is an important valuation of economic value based on established positions; maintaining the status quo. While some may prefer to refer to it as a principle of allocation, we will keep calling it a justice principle, for simplicity and readability.  

\section{Research design}
\label{sec:method}
\subsection{highRES modelling framework}
For this modelling exercise, we use the open-source high spatial and temporal Resolution Electricity System model for Europe (highRES) \cite{priceHighRESEuropeHighSpatial2022}. highRES minimises the total system cost (operating and annualised investment costs) by optimising the dispatch and spatial allocating of capacity investment of power plants, storage technologies and transmission grid extension. With highRES we generate snapshots of different electricity system designs for an interconnected European electricity system in 2050. The model version which we apply in this case consists of 28 nodes, each representing one of the EU27 (excluding Cyprus and Malta) + Norway, The United Kingdom and Switzerland. The model is a hybrid greenfield model, which does not consider existing infrastructure, except for hydropower reservoirs and capacities, which remain fixed at the current state of the system. A detailed overview of core model assumptions can be found in appendix \ref{app:model_desc} and the associated GitHub repository\footnote{https://github.com/OskarVagero/MENOFS}, which contains the model formulation, data necessary to replicate both modelling and analysis, as well as descriptions of major changes from previously published versions \cite{priceHighRESEuropeHighSpatial2022,priceRoleNewNuclear2023}.

\subsection{Modelling to Generate Alternatives (MGA)}
To study how the distributional justice of an electricity system changes with how justice is defined, we use the Modelling to Generate Alternatives (MGA) technique to find a variety of system configurations which can be quantitatively evaluated in terms of their distributional justice (see section \ref{sec:gini_coef}). The MGA technique is applied to sample alternative system configurations that are near-optimal in the sense that they only deviate from the cost-optimal model solution within a tolerable degree, while optimising for a new objective. 

It is important to state that MGA is not the only technique one could apply to study distributional justice, as imposing equity related constraints \cite{neumannCostsRegionalEquity2021} or multi-objective optimisation \cite{drechslerEfficientEquitableSpatial2017} is featured in the literature. However, given that the focus of this work is to discuss and reflect on current and future modelling practices and the implications of implicit assumptions, we chose MGA as it has acquired traction in energy systems analysis \cite{vageroCanWeOptimise2023,lombardiWhatRedundantWhat2023}. We believe that this allows for more comparison with the existing literature and makes the results more relevant for future modelling practices. 

For this work, we utilise what Pedersen et al. \cite{pedersenModelingAllAlternative2021} call ‘One-At-the-Time (OAT) approach, previously applied by Refs. \cite{neumannNearoptimalFeasibleSpace2021,neumannBroadRangesInvestment2023}, and the process can be summarised as follows: 

\begin{enumerate}
    \item A baseline model scenario is generated by running the model in its normal function, minimising the total system cost, resulting in a cost-optimal objective value $f(x^*)$.
    \item An allowable cost-increase $\epsilon$ (slack) to the cost-optimal total system cost is introduced, which forms a new constraint, allowing the total system cost of subsequent model runs to be $f(x) = f(x^*) \cdot (1 + \epsilon)$. 
    \item For each MGA scenario, a new objective is introduced, which for this OAT approach revolve around min- and maximising the installed capacity of various electricity generation technologies, for each country node individually as well as aggregated. We have chosen to limit ourselves to only onshore- and offshore wind, solar and nuclear, as e.g. hydropower is not expandable and gas is heavily constrained by emission limits. 
\end{enumerate}

Following this approach, we generate 696 MGA scenarios, maximising and minimising the installed capacity of onshore- and offshore wind, solar and nuclear for each of the 28 individual countries and for the system as a whole, at three discrete slack levels, 5, 10 and 15\%. These slack levels are consistent with previous studies \cite{sasseDistributionalTradeoffsRegionally2019,neumannNearoptimalFeasibleSpace2021,neumannBroadRangesInvestment2023,grochowiczIntersectingNearoptimalSpaces2023} and within the range of how far from cost-optimal real-world energy transitions have been found to be \cite{trutnevyteDoesCostOptimization2016}.

To exemplify the process, we can imagine that the baseline model solve optimally and that the total system cost $f(x^*) = 100$. For two subsequent MGA scenarios, where $\epsilon = 0.05$, the model will be constrained to a total system cost of $f(x^*) \cdot (1 + \epsilon) = 105$ and tasked with new objectives. If the objective is to respectively maximise and minimise the onshore wind deployment in France, the model will invest less/more (relative to the cost-optimising scenario) in alternative technologies across the system, to maximise/minimise the room for onshore wind, while ensuring that all the normal constraints are met. 

With the OAT approach, we sample only the extreme points at the vertices of the solution space, meaning that there are interior points which may be equally valid system configurations, but which we overlook, see e.g. Neumann and Brown \cite[Fig.~1]{neumannNearoptimalFeasibleSpace2021} for a simplified graphical illustration of the multidimensional near-optimal feasible space. Although there are more sophisticated methods of sampling the near-optimal feasible space (see e.g. \cite{lombardiPolicyDecisionSupport2020,pedersenModelingAllAlternative2021,grochowiczIntersectingNearoptimalSpaces2023}), it is not our ambition to acquire a full representation from our sampling, as the purpose of our work is to illustrate the central point around different interpretations and definitions of justice. Furthermore, the OAT approach is useful to understand the structure of the decision space \cite{neumannNearoptimalFeasibleSpace2021}. However, it also generates system configurations which are nonsensical from a practical point of view. In our sample, this is particularly true when the model is tasked to maximise the installed capacity of some technology, as it may install capacity which is then not used to generate electricity (as that is not an objective of the model). This work is meant to be illustrative, and by including the maximising scenarios, we get a larger diversity in electricity systems, even if they may not be implemented in reality.  

\subsection{Gini-coefficient}
\label{sec:gini_coef}
The Gini index or Gini coefficient is a statistical indicator for studying the variability of objects. In the original publication, the Gini coefficient is defined as: "the mean difference from all observed quantities" \cite{cerianiOriginsGiniIndex2012}. It is most commonly applied as a way to measure income inequality, but has been applied as a statistical indicator in a wide range of fields, such as energy systems analysis \cite{sasseDistributionalTradeoffsRegionally2019,sasseRegionalImpactsElectricity2020,neumannNearoptimalFeasibleSpace2021} epidemiology \cite{abelesGiniCoefficientUseful2020} and variation in grain yield \cite{sadrasUseLorenzCurves2004}.

One way of mathematically defining the Gini coefficient, $G$ is:
\begin{equation}
\label{eqn:gini}
    G = \frac{\sum_{i=1}^{n}\sum_{j=1}^{n}\abs{x_i-x_j}}{2\sum_{i=1}^{n}\sum_{j=1}^{n}x_j} = \frac{\sum_{i=1}^{n}\sum_{j=1}^{n}\abs{x_i-x_j}}{2n^2\bar{x}},
\end{equation}

where $x_i$ is the studied factor (to be distributed) of study object $i$ and there are $n$ study objects. $x_j$ represents a different study object compared to the focal one ($i$), and $\bar{x}$ is the arithmetic mean value of $x_i$. Applied to an electricity system model, the studied factor ($x$) could be e.g. installed generation capacity or electricity generated in a country or region ($i$). Assigning each country or region the same weight largely skews the results given that the population, land-area available and electricity consumption widely differs between the studied objects. As such, it is possible to weigh the studied factor by instead considering e.g. installed generation capacity per capita, consumption or land area (see section \ref{sec:just_def}).

Since the relationship between the different variables in equation \ref{eqn:gini} are non-linear, we are unable to include these calculations within our linear optimisation model. As such, we use modelling to generate alternatives (MGA) (previously described in section 3.2) to sample the near-optimal feasible solution space, and generate diverse electricity system configurations. During the analysis and post-processing of the results, we calculate the Gini coefficients for each principle of distributive justice and MGA scenario, resulting in 4,176 different values of the Gini coefficient. 

\subsection{Justice definitions}
\label{sec:just_def}
In addition to describing the theories supporting the principles of distribution (introduced in section \ref{sec:theory}) and defining the Gini coefficient used to evaluate the equity of a system, we need to operationally define the justice principles. The contribution of different regions changes with the definition of the burden or benefit distributed ($x_i$). Additional to determining the appropriate definition of subject of moral worth (i.e. who deserves moral consideration), one also needs to consider what Sasse \& Trutnevyte \cite{sasseDistributionalTradeoffsRegionally2019} call equity factors (EF), and what it is that should be equally distributed. Previous modelling studies have considered, among other things, the distribution of electricity generation infrastructure (e.g. wind turbines or solar panels) \cite{drechslerEfficientEquitableSpatial2017}, emission-rights \cite{pedersenUsingModelingAll2023} or energy-associated economic impacts (e.g. job opportunities) \cite{sasseRegionalImpactsElectricity2020}.  

For the efficiency principle, we conform to conventional cost-optimising modelling, which simply requires designing the system while minimising the total system cost. As such, there is nothing to distribute within the system, and the Gini coefficient is inapplicable. 

Operationalising the equality, capability, and historical responsibility principles all follow similar logic. We consider installed capacity of electricity generating infrastructure to be the equity factor that should be distributed. The rationale behind distributing electricity generation capacity is based on the externalities and impacts that come with them, not only where the investments stem from. For example, wind power is often considered to have an impact on the visual aesthetic of a location \cite{tsaniOutSightOut2024,tsaniQuantifyingSocialFactors2024}, so even if e.g. private companies invest in a wind power plant, the local community will end up with the visual impact. Similarly, generation capacity may also lead to employment opportunities, which will positively impact the local community. As this work is meant to be illustrative, we use the generation capacity as a proxy for these impacts, which creates a balance compared to e.g. focusing on only the job opportunities. The studied factor ($x_i$) is then the fraction of the installed capacity, weighted by population, purchasing power parity (PPP) adjusted GDP or cumulative historic CO\textsubscript{2} emissions, which is presented in equation 2 through 4:

\begin{align}
    x_i = \frac{P_i}{\rho_i}, \forall i \in C \\
    x_i = \frac{P_i}{E_i}, \forall i \in C \\
    x_i = \frac{P_i}{CC_i}, \forall i \in C
\end{align}

with:\newline
    \-\ \hspace{20pt} $P_i$ \hspace{20pt} installed capacity \\
    \-\ \hspace{20pt} $\rho_i$ \hspace{20pt} population \\
    \-\ \hspace{20pt} $E_i$ \hspace{20pt} PPP adjusted GDP \\
    \-\ \hspace{20pt} $CC_i$ \hspace{20pt} cumulative CO\textsubscript{2} emitted between 1900-2021 \\
    \-\ \hspace{20pt} $C$ \hspace{20pt} Set of countries \\

For the self-sufficiency principle, the installed capacity is less relevant than the amount of electricity generated, and we as such change the equity factor to be total electricity generated in a year, which is weighted by the annual electricity demand, as seen in equation 5:

\begin{equation}
    x_i = \frac{G_i}{D_i}, \forall i \in C
\end{equation}

where, $G_i$ is total annual electricity generation and $D_i$ is the annual electricity demand.

The land burden principle focuses on the land-use conflicts that electricity generation infrastructure can have with non-modelled aspects, such as agricultural or recreational land, or other alternative uses. It connects societal and ecological burden from electricity generation infrastructure to land area occupied and population density through the indicator burden level, originally described by Christ et al. \cite{christWindEnergyScenarios2017} and presented in modified form in equation 6:

\begin{equation}
    x_i = \frac{A_{occupied,i}}{A_{avail,i}}, \forall i \in C
\end{equation}

where $A_{occupied,i}$ is the area occupied by energy generating technologies, $A_{avail,i}$ is the area available. Installed capacity of electricity generating infrastructure is implicitly considered, but also weighed based on how much area is occupied by different technologies. We base the area factors on Chen et al. \cite[Tbl.~A6]{chenBalancingGHGMitigation2022}, where solar PV and onshore wind occupy a considerably larger area than any other technology. As such, a system is distributively just (according to this specific interpretation) when the same share of a country’s buildable area, is occupied by electricity generation infrastructure. This does not result in the same area available for everyone, only that everyone ‘sacrifices’ the same share of their available resources.

Lastly, for the grandfathering principle, we assess the change from status quo, meaning that if each country keep the same share of the total installed capacity, the system will be perfectly equitable. The equity factor is as such:

\begin{equation}
    x_i = \frac{P_i}{P_{old}}, \forall i \in C
\end{equation}

where $P_{old,i}$, is the installed capacity already built in 2022. As such, an equitable system should strive to reduce the impact on already established value-chains, maintaining the same shares of capacities as the current electricity system. 

\begin{table*}[width=\textwidth,cols=5,pos=h!]
  \caption{Overview of justice principles}
  \begin{tabular*}{\tblwidth}{@{} p{2.0cm}p{4.0cm}p{4.5cm}p{2cm}p{2cm}@{}}
    \toprule
    Principle   & Description & Example of Equity factor(s) & Denominators & Reference \\
    \midrule
    Efficiency  & Distribute to minimise total system cost & None & None &  \\
    Equality    & Distribute equally & Installed capacity / investment costs / emissions & Population & \cite{drechslerEfficientEquitableSpatial2017,sasseDistributionalTradeoffsRegionally2019,pedersenModelingAllAlternative2021,neumannNearoptimalFeasibleSpace2021}  \\
    Self-sufficiency & Distribute for national independency & Electricity generation & Demand  & \cite{neumannCostsRegionalEquity2021,trondleTradeOffsGeographicScale2020,hohneRegionalGHGReduction2014} \\
    Capability & Distribute based on socio-economic capacity & Installed capacity / investment costs / emissions & PPP adjusted GDP & \cite{pedersenModelingAllAlternative2021,hohneRegionalGHGReduction2014} \\
    Historic responsibility & Distribute so that historical beneficiaries do more & Installed capacity / investment costs / emissions & Per capita CO\textsubscript{2} emitted since 1900 & \cite{pedersenModelingAllAlternative2021,hohneRegionalGHGReduction2014} \\
    Land burden & Distribute for an equal share of available area & Occupied area  & Available area & \cite{chenBalancingGHGMitigation2022,christWindEnergyScenarios2017} \\
    Grandfathering & Distribute to reduce the impact on established value-chains & Installed capacity / investment costs / emissions & Current system capacities & \cite{zhouCarbonDioxideEmissions2016,pedersenModelingAllAlternative2021} \\
   \bottomrule
  \end{tabular*}
  \label{tbl:principles}
\end{table*}

The denominators used in the calculation of the Gini coefficient, shown in table \ref{tbl:principles} are based on a mix of descriptive statistics (population, GDP, emissions and currently installed capacity) and modelling assumptions (available land area and electricity demand). All descriptive statistics are for 2022, which is the last year of when there is available data across all factors. Population data is from Eurostat \cite{eurostatPopulationChangeDemographic2022}, PPP GDP from the World Bank \cite{worldbankGDPPPPCurrent2023}, cumulative CO\textsubscript{2} emissions from the Global Carbon Budget \cite{friedlingsteinGlobalCarbonBudget2022} and installed energy generation capacities from ENTSO-E Transparency Portal \cite{entso-eENTSOETransparencyPlatform2022}. Demand and land-use data is presented in appendix \ref{app:model_desc}. An overview of the spatial distribution of the descriptive statistics is presented in appendix \ref{app:desc_stat} and figure \ref{fig:heat_stats}.

\section{Results}
\label{sec:results}
Before we explore the alternative system configurations generated with the MGA technique, we can look at distributional aspects of the 2022 version of the real European electricity system as well as the cost-optimal model scenario generated with highRES. This gives us an overview of the starting point of the system, some initial insight into how the relevant indicators are distributed across the system, and an assessment of these two systems' distributional justice according to the principles defined in section \ref{sec:theory}. Data for the system in 2022 is sampled from ENTSO-E \cite{entso-eENTSOETransparencyPlatform2022}. 

\begin{figure}[h!]
    \centering
    \includegraphics[width=\textwidth]{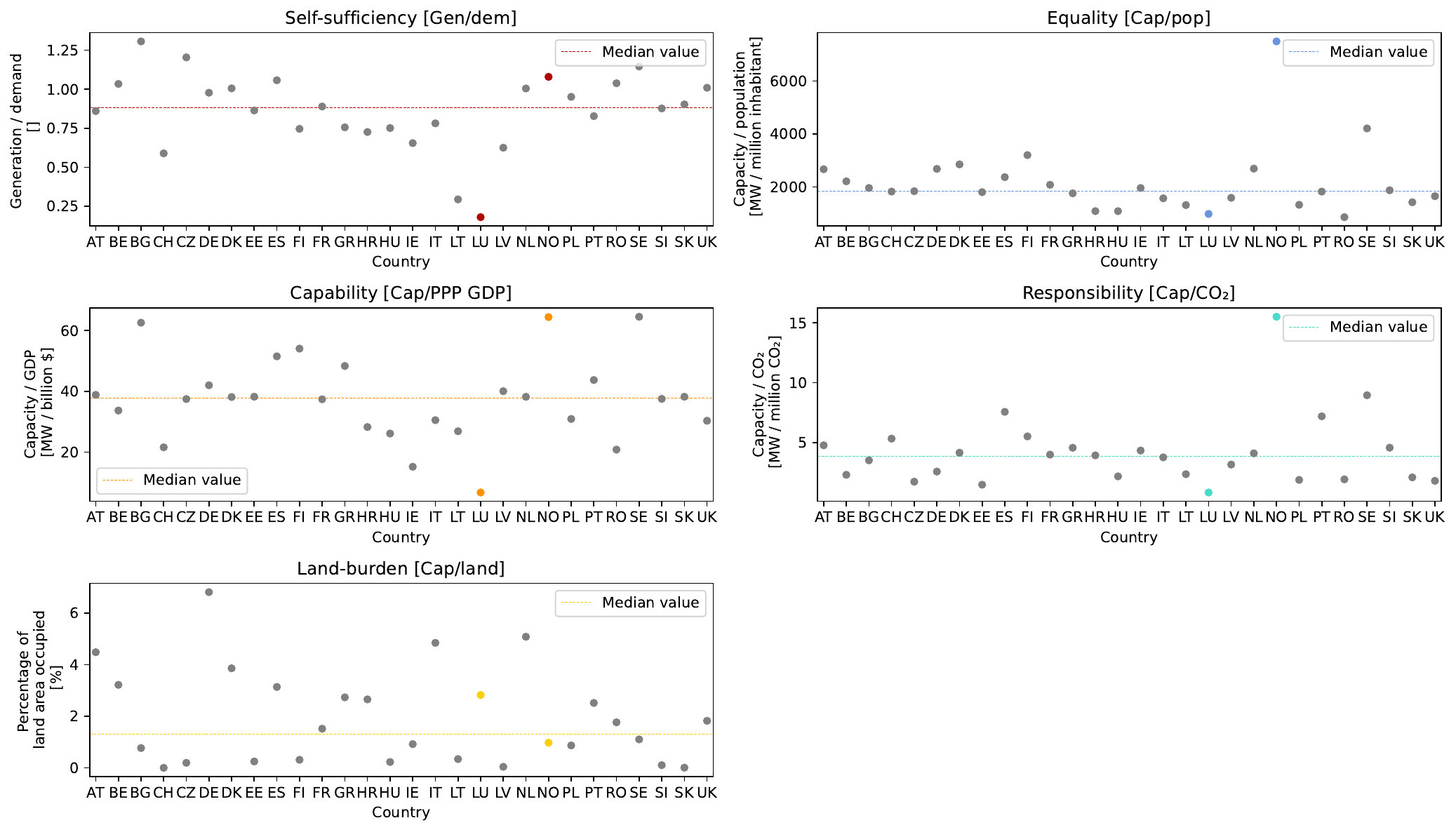}
    \caption{Equity factor ($x_i$) values for each country $i$, according to different distributional justice principles, based on the electricity system in 2022. NO and LU in colour, for illustrative purposes}
    \label{fig:current_system}
\end{figure}

To begin with, figure \ref{fig:current_system} shows the performance of each individual country according to five of the different justice principles outlined in section \ref{sec:theory} and their relevant equity factor. The grandfathering principle is excluded in this figure, as it is only applicable in the MGA scenarios. Here we can easily see that the performance of countries, relative to one another, changes based on what quantitative measure is used in the assessment. Luxembourg (LU), a small country which currently relies on imported electricity from neighbouring countries to supply about 80\% of its electricity demand \cite{entso-eENTSOETransparencyPlatform2022}, performs poorly according to the first four assessments, with either the lowest or second-lowest value. However, when it comes to land area, Luxembourg uses a high share of its available land area relative to other countries. For Norway (NO) the situation is reversed, with high scores in all measures, except for when one assesses the share of land area occupied, due to its low population density. 

\begin{figure}
    \centering
    \includegraphics[width=0.9\textwidth]{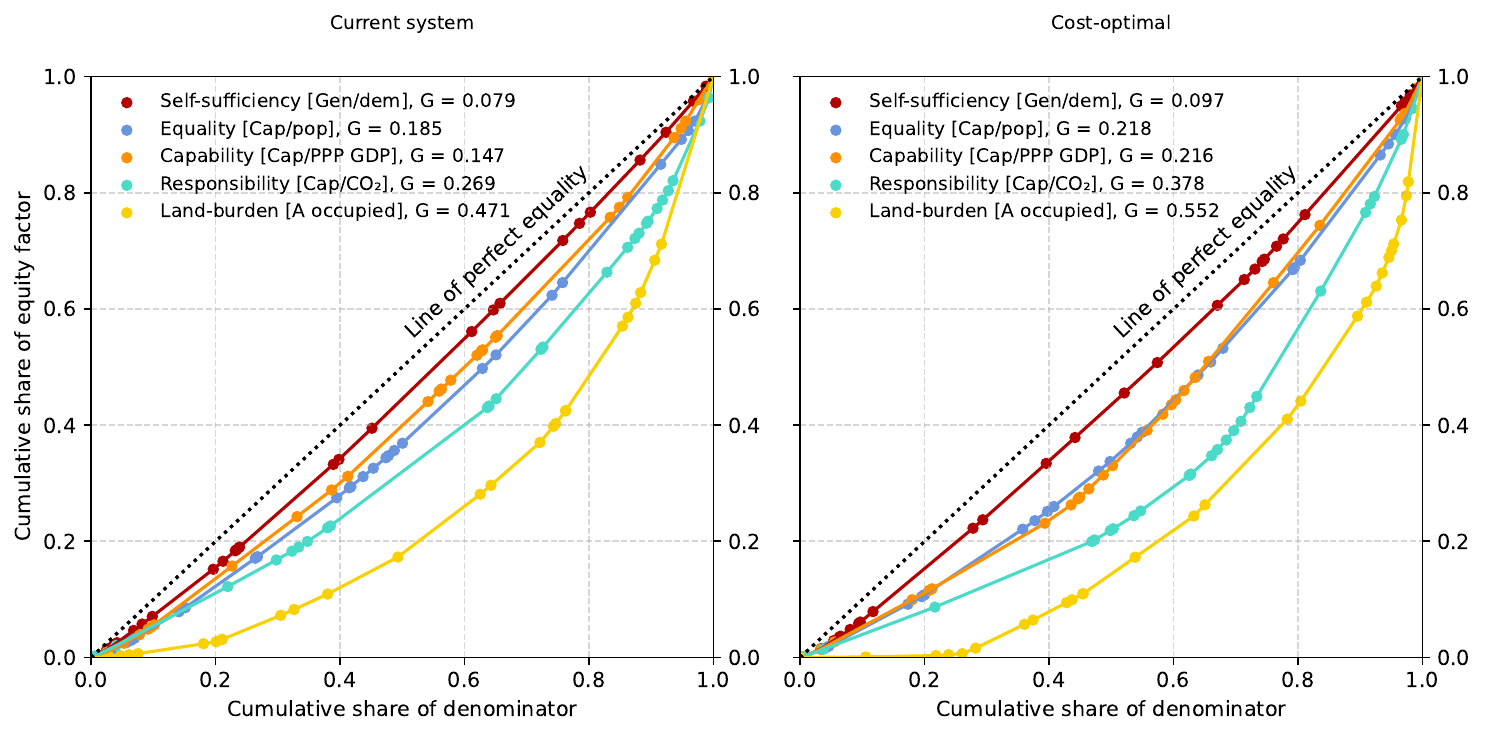}
    \caption{Lorenz curves | Resulting Gini coefficients for the current system and cost-optimal scenario. Values close to 0 indicate high equity, whereas values close to 1 indicate high inequity. Dots are individual data points.}
    \label{fig:lorenz}
\end{figure}

The same can also be shown from calculating the Gini coefficient (see section \ref{sec:gini_coef}) as an assessment of the distributional equity for the system as a whole. According to the self-sufficiency principle, the system has a rather high distributional equity (G = 0.079 and 0.097), but a high inequity for the land-burden principle (G = 0.483 and 0.552), both for the current system and the cost-optimal scenario. The other principles lie somewhere in between, as can be seen in figure \ref{fig:lorenz}. Notably, the Gini coefficient is also higher in the cost-optimal scenario than in the current system, for all five justice principles which can be compared. 

\subsection{Gini coefficients of future electricity systems}

\begin{figure}
    \centering
    \includegraphics[width=\textwidth]{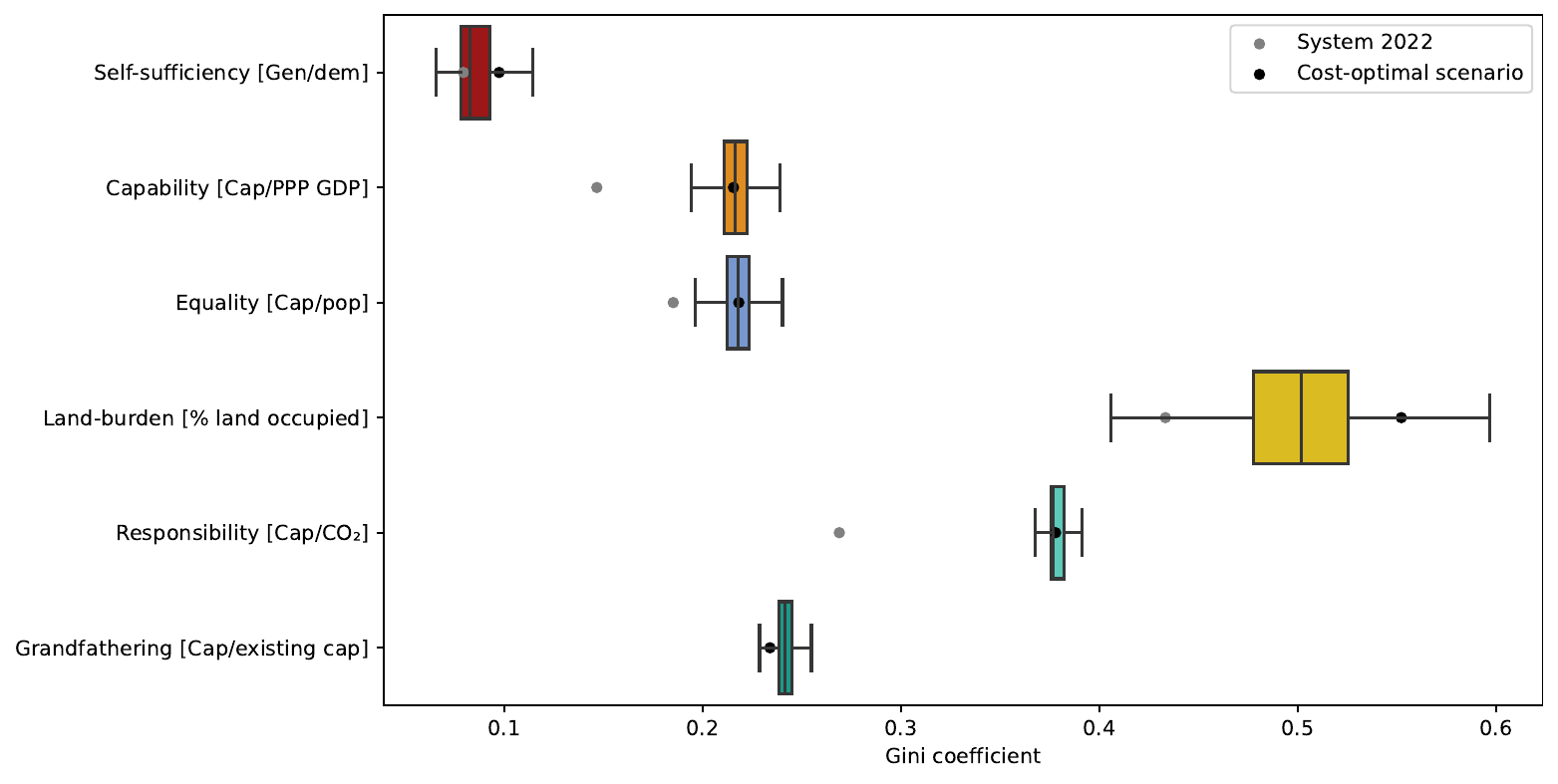}
    \caption{Box plot of the resulting 4176 Gini coefficients for the scenarios as well as individual points for the cost-optimal scenario and system in 2022. Whiskers represent 1.5×IQR, and outliers are not shown in the figure.}
    \label{fig:boxplot}
\end{figure}

Furthermore, in figure \ref{fig:boxplot}, we can see the distribution of the Gini coefficient for each definition of distributional equity. The statistical dispersion of the Gini coefficients is summarised in table \ref{tbl:desc_stat} in the appendix. From figure \ref{fig:boxplot} and table \ref{tbl:desc_stat}, we can draw some general conclusions around how the Gini coefficient is distributed, such as that the self-sufficiency principle has the smallest range and lowest standard deviation. For the other principles, the dispersion is greater, although in various ways. The responsibility principle has a relatively large range between the highest and the lowest value, but a lower IQR than the remaining three principles. Furthermore, the resulting Gini coefficient of the different distributional justice principles are largely different. Only the capability and equality principles have similar values. The scenarios contain the most distributional inequity according to the land-burden principle, followed by the responsibility principle, with the self-sufficiency principle showing the opposite. The choice of principle therefore has a large impact on the evaluation of how much distributional inequity is in a system. 

We can further identify the MGA scenarios with the best (lowest) Gini coefficient according to each justice principle, which is illustrated in figure \ref{fig:top_perf} in appendix \ref{app:desc_stat}. As such, out of all MGA scenarios, maximising the installed capacity of solar power in Germany, with a 10\% slack (DEMAXSOL10), results in the least inequity, according to the self-sufficiency principle. However, if equity is expressed through the capability principle, maximising onshore wind power in Germany with a slack of 5\% (DEMAXON05) results in the least inequity. Furthermore, the trade-off between cost-efficiency and improving equity is different for the different justice principles. For example, with the capability principle, we find an improvement by 21\% (4.6 ppt) at a 15\% total system cost increase and for the land-burden principle a 14\% (7.1 ppt) for the same cost increase. 

\begin{figure}
    \centering
    \includegraphics[width=\textwidth]{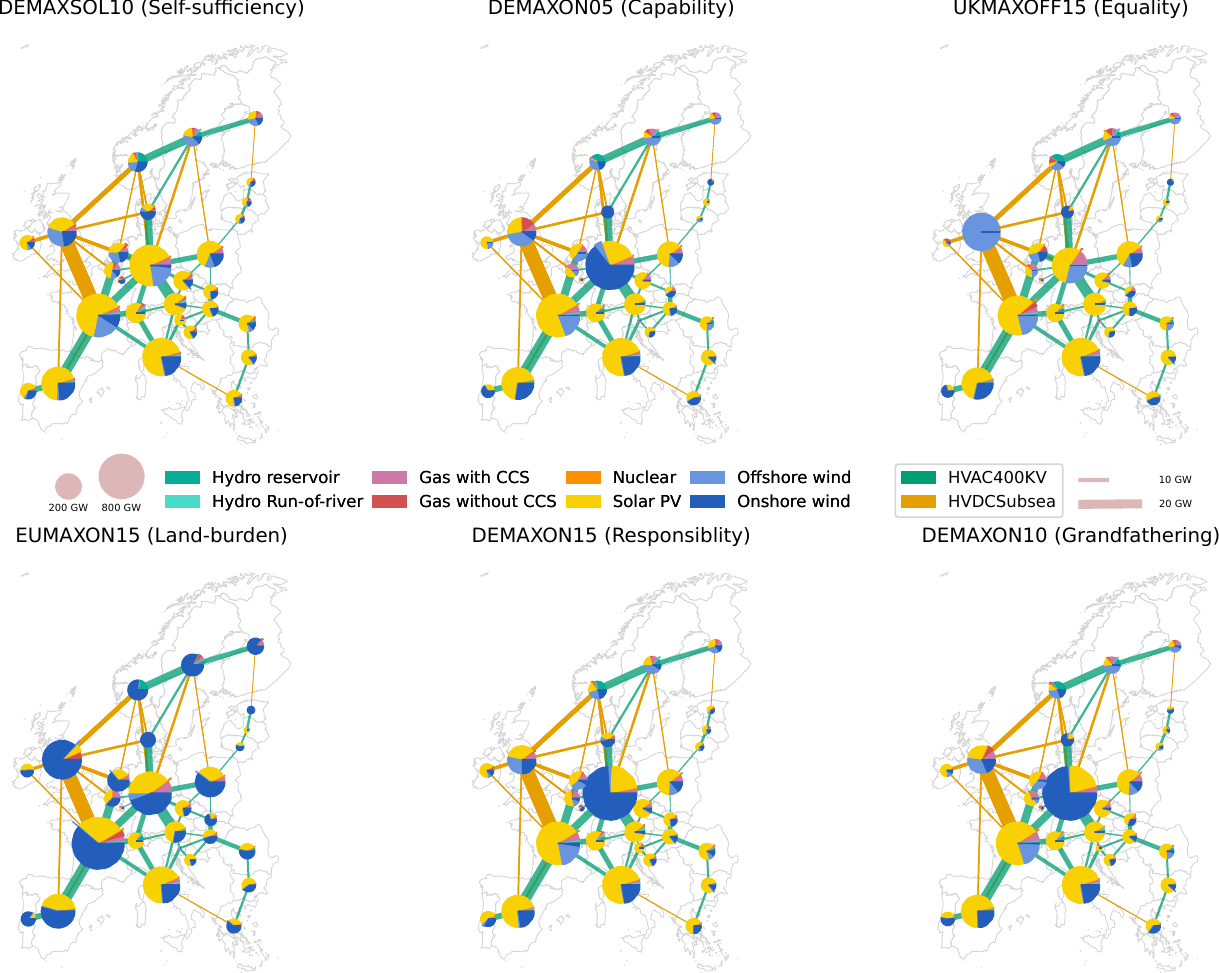}
    \caption{Allocation of electricity infrastructure in the top performing (lowest Gini coefficient) MGA scenarios, according to the six different distributional justice principles.}
    \label{fig:top_perf_map}
\end{figure}

Looking more in-depth at the system design for the top performing MGA scenarios, we can study how the model decides to allocate electricity generating infrastructure, as illustrated in figure \ref{fig:top_perf_map}. The pie-chart represent the share of installed generation capacity in a country, with the size representative of the total amount. Transmission infrastructure is shown through the orange and green lines, which represent subsea cables (HVDC) and overhead transmission lines (HVAC). Naturally, the optimisation objective of each specific MGA scenario plays a big role in the allocation, but it is not the only aspect which changes. For example, in DEMAXSOL10 there is an additional 122 GW of installed onshore wind capacity across the system, and 16 GW less natural gas to facilitate more solar capacity. In DEMAXON15, the system invests in 262 GW of additional solar capacity (and 128 GW less offshore wind) in addition to changes in onshore wind. Although the details of the figures are difficult to discern, they are only meant to be illustrative of how the system configurations are different. As such, the importance is not on the details and whether Germany builds 400 or 200 GW of solar, as in DEMAXSOL10 and DEMAXON05, but rather that the most distributionally just system is noticeable different depending on which perspective one takes (and which principle is used). The six systems represent system designs with the least distributional inequity (lowest Gini coefficient) based on which of the justice principles is applied. 

\section{Discussion}
\label{sec:discussion}
First and foremost, we show that the application of different distributional justice principles in energy system optimisation modelling is feasible and that the traditional approach - maximisation of efficient distribution of energy technology at the lowest cost possible - is one among a number of alternatives. As such, this is an important finding, as this key for distribution has a close to paradigmatic standing in economic analyses and techno-economic modelling. At the same time, we show that it is far from value-neutral. This makes investigating alternatives important in and of itself, so justice-related outcomes can be made more transparent.

Furthermore, the results clearly indicate that the distributive outcomes of the different principles are sufficiently different to be regarded as important outputs for further discussion. Already from studying the electricity system as it is today (figure \ref{fig:current_system} and \ref{fig:lorenz}), and further strengthened by the spread of Gini coefficients in figure \ref{fig:boxplot}, we find that the evaluation of distributional equity changes with the application of different justice principles, both in terms of internal ranking of countries and the overall resulting Gini coefficient. While this finding is very intuitive and would not require any modelling, the particular application in energy systems optimisation modelling is important. Given the increasing focus on studying justice aspects in energy systems optimisation modelling, and the limited reflection on this particular normative assumption in the field, this needs to be brought to the attention of modellers. 

Overall, our results show the anticipated spread of Gini coefficients, with the best (lowest) value ranging from 0.064 to 0.386, based on the different definitions we have applied. The design of the “justice-optimal” energy system varies considerably, on the basis of how distributional justice is interpreted. The results also show that the self-sufficiency- and equality principles, which have been applied most frequently in earlier works \cite{vageroCanWeOptimise2023}, are also the ones with the lowest Gini coefficient, both in the cost-optimal model run and for the top-performing MGA scenarios. Analysing specifically these justice principles may as such result in more modest assessments of how just/unjust an electricity system is perceived to be, as well as the potential for improvement. 

Similar to previous research \cite{neumannCostsRegionalEquity2021}, we find that the cost-optimal future electricity system tends to lead to higher levels of inequity, not only when defined in terms of self-sufficiency, but also for the distribution of electricity generation capacities per inhabitant, economic capability, cumulative historic emissions and land-area availability. Noticeably, for the capacity and responsibility principles, neither of the scenarios show a lower inequity (Gini coefficient) than the system in 2022, even with up to a 15 percent cost-increase. This reflects how modelling results often deviate from existing electricity systems. As is known from real-life political situations, uncritically applying cost-optimised ideal results may therefore be politically difficult, and compensations or adjustments can be expected in actual energy system application \cite{inderbergIdentifyingAnalysingImportant2024}.

Exploring alternative system configurations and evaluating them based on different principles of distributional justice allows for exploring how the trade-off between cost-efficiency and distributional equity changes with how equity is defined. While the results of Sasse \& Trutnevyte \cite{sasseDistributionalTradeoffsRegionally2019} showed a similar trade-off for both the equality and self-sufficiency principle for a 2035 Swiss electricity system, including a wider interpretation of distributive justice leads to a larger spread in the results. The equity improvement of the top performing scenarios relative to the cost-optimal scenario range from 2.4 to 7.1\%pt., and 14.4 to 27.5\%. Previous studies have also shown how the spatial allocation changes between a cost-optimal scenario and more equitable system designs. Both Drechsler \cite{drechslerEfficientEquitableSpatial2017} and Sasse \& Trutnevyte \cite{sasseDistributionalTradeoffsRegionally2019} noticed a shift from onshore wind to open-field solar PV in the more equitable electricity system (based on a self-sufficiency and/or equality principle) for Germany and Switzerland, respectively. Neumann and Brown \cite{neumannNearoptimalFeasibleSpace2021} showed that transmission network expansion can lead to increased inequity (when considering equality-based justice), while investing in storage can increase equity somewhat (although not in a zero-emission system). Our continental-scale results do not indicate that specific technologies are required for more equitable electricity systems, as it depends more on the interpretation of justice principle applied. This could, however, be impacted by aspects such as our sampling of the near-optimal feasible space, spatial- and temporal resolution of the modelling setup and assumptions of e.g. land-use intensity of technologies.

Justice is getting more political traction also when thinking of the necessary speed of energy transitions \cite{newellNavigatingTensionsRapid2022}, with social opposition due to perceived inequities being a major hindrance \cite{eikelandWindChangeNorway2023}. Utilising model results for policy and decision-making further elevates the need for openness and explicit assumptions, as this research represents an initial step in the direction of. It is important to ensure that model results are not used for legitimating ‘hidden morality’ \cite{vanuffelenDetectingEnergyInjustices2023} political and clearly justice related decisions under the auspices of being value-free or neutral. Decisions should be transparently based on explicit assumptions rather than on concealed moral biases disguised as neutral, such as that the efficiency principle would be value-free. A better understanding of the distributive justice implications of cost-optimal modelling results may elevate the discussion on how to interpret, and utilise, modelling results. Additionally, open modelling builds trust and legitimacy in results and processes \cite{susserModelbasedPolicymakingPolicybased2021,pfenningerImportanceOpenData2017}, which is important for the acceptance and adoption of energy policy. However, it is not necessarily only a matter of providing open data, open source-code and publishing open access, but modelling practices may need adaptations too, if the results are to be perceived as truly accessible.

The wide diversity of distributional inequity shown in our results, which are limited to six different justice principles, further indicates that selecting a single principle may not be justifiable. This aligns with, for example, Drechsler et al. \cite{drechslerEfficientEquitableSpatial2017} who saw a disagreement in which spatial allocation was considered fair, based on survey results and choice experiments. The combination of principles and clearly showing the implications of the alternatives are in our view likely to lead to a better-informed menu of options for decision-makers, to make important choices on behalf of energy system developments and implications for society. This is currently not the state of things. It is also possible to involve stakeholders and decision-makers in several phases of the modelling process \cite{mcgookinAdvancingParticipatoryEnergy2024}, with different justice principles co-defined from early on, or by providing a set of model results that stakeholders can engage with and select between \cite{pickeringDiversityOptionsEliminate2022}. In essence, insufficient consideration of definitions and related assumptions in modelling can lead to results with false social justice claims, which nevertheless hold hidden consequences for particular societal groups. 

While previous studies have contributed with valuable insights around the nature of distributional inequity, and proven its relevance in energy systems, we demonstrate the importance of being explicit about equity definitions and justice principles in modelling. However, these are to be considered examples, and we do not claim that the definitions applied in this work are exhaustive or representative of interpretations of equitable distributions. As such, these results only highlight differences, and are only intended to illuminate  challenges with narrow interpretations of often implicit justice claims in modelling and beyond. Furthermore, the Gini coefficient only measures the equity of a system as a whole, and not the distribution of inequity within systems. There are different distributions for the same Gini coefficient. Similarly, when applied on a country-level, it does not account for regional disparities. The Gini coefficient is a frequent statistical indicator for measuring uniformity of a distribution, but is not necessarily the only or best indicator to quantitatively assess distributional justice. Our choice of using the Gini coefficient was based on previous research in energy systems modelling, which our work was a response to, however further research in this and other quantitative fields should consider whether there are other indicators that may give other insights and how they compare or could be combined to get a more detailed understanding of the system. 

Choosing the right, or relevant, equity factor is also important. Studying the distribution of electricity generation infrastructure, as the equity factor in this work, represents an approach which tries to balance several impacts (positive and negative) of different technologies. If one only care for a specific impact, like landscape visual impact, or employment opportunities, that should be studied directly, but the problem is often more complex. Assessing different equity factors adds nuances to the distributional justice principles, and the results are likely different if it is employment or visual impact that should be shared e.g. equally on a per capita basis (in the example of the equality principles). Nevertheless, the importance of being explicit and sharing assumptions remains a crucial point.

Finally, this study, along with a significant amount of research combining energy systems analysis and justice, almost exclusively focuses on the distributional end-state of the system and not on the processes and institutional context leading us there \cite{vageroCanWeOptimise2023}. Young, for example \cite[Ch.~1]{youngJusticePoliticsDifference2011}, challenges and criticises the so-called distributive paradigm in philosophical theories of justice. While distribution is clearly important, there is a need for considering “[...] a wider context that also includes actions, decisions about action, and provision of the means to develop and exercise capacities” \cite[p.~16]{youngJusticePoliticsDifference2011}. More recently, we can relate this to the energy justice framework \cite{jenkinsEnergyJusticeConceptual2016} and the dimensions of the framework relating to procedural and recognition aspects of justice, as well as restorative justice that addresses remediation of previous injustices. This could for example cater to meaningful participation and inclusion in decision processes, or co-decisions in licensing of large energy infrastructure. In the context of energy systems optimisation modelling, the strong focus on quantitative distributional justice along with the challenge of normative assumptions, as outlined in this work, indicate that engagement between modellers and stakeholders, e.g. through participatory modelling process \cite{mcgookinAdvancingParticipatoryEnergy2024}, could ameliorate some of the limitations of modelling. Recently, a new MGA technique, called human-in-the-loop (HITL) MGA have been proposed to include stakeholders' system design preferences into the MGA workflow \cite{lombardiHumanloopMGAGenerate2024}, allowing for a more guided search of system configurations.

\section{Conclusion}
\label{sec:conclusion}
In this article, we argue that justice perspectives are grossly under-addressed, and simplified, in energy systems optimisation modelling studies and practices. To take some initial steps to address this, we have determined six applicable distributive justice principles to apply to energy modelling, while exploring alternative systems than only the cost-optimal efficient distribution. We have used an electricity system optimisation model, and the modelling to generate alternatives (MGA) technique to illuminate the implications of applying different definitions of justice (i.e. justice principles) to the assessment of distributional equity in future European electricity systems. We generated 696 different electricity system designs and evaluated their distributional equity from the perspective of six different justice principles. 

From our analysis, we found that the assessment of justice in electricity systems is highly dependent on how distributional justice is perceived and framed by modellers. This is important, as recent policy initiatives such as the European Green Deal not only  focus on transforming  future energy systems to be net-zero but also just. It is thus crucial to discuss how the choice of justice definition affects the model results and subsequent policy decisions. 

While previous research has introduced questions of just energy systems particularly concerning distributional aspects to  the academic literature, this work shows that applying a significantly wider scope of justice principles is needed. While our included justice principles are far from exhaustive, this work represents some key initial steps for this research literature, within modelling and beyond. We therefore emphasise the need for further and explicit discussions on normative uncertainty and how assumptions may impact results, conclusions and policy recommendation, particularly within scholarly modelling practices. 

We should remind ourselves that justice in energy systems and energy transitions are not limited to the distributional end-state of a system. While material distribution is fundamental, processes of decision-making and involvement as well as recognition of different needs, views, and epistemologies may be considered equally important. Through processes of collaboration and involvement, there may also be ways to better manage  normative uncertainty and the processes of attributing justice within energy system scenarios and decision-making. In the context of energy systems optimisation modelling, participatory processes is one option with the potential to ameliorate several of the limitations stated, by providing study-specific context to the construction and justification of justice principles.

\appendix
\section{Model description}
\label{app:model_desc}
highRES is a linear cost-optimising electricity system model, designed to specifically analyse electricity systems with a high level of variable renewable energy sources. The model minimises electricity system costs (operating costs and annualised investment costs) to meet hourly demand subject to a number of technical constraints; thereby optimising the dispatch and locational investment into power plants, storage and transmission grid extension. Some aspects, such as the availability of materials, and build-rates for the different capacity investments are not included in the model. The modelling framework is described in Price and Zeyringer \cite{priceHighRESEuropeHighSpatial2022} and Price et al. \cite{priceRoleNewNuclear2023} marks the most recent publication using highRES in a European context.

The model balances supply and demand at an hourly resolution for all nodes in the model. Demand time series are originally based on historical data from the European Network of Transmission System Operators for Electricity (ENTSO-E) Transparency Platform \cite{entso-eENTSOETransparencyPlatform2022}, but need to be adjusted to account for inconsistencies and missing data. This has previously been done by van der Most et al. \cite{vandermostExtremeEventsEuropean2022}, who used climate data and applied a logistic smooth transmission regression (LSTR) model to the ENTSO-E dataset to correlate historical electricity demand to temperature and generate daily electricity demand for a set of European countries. Subsequently, Frysztacki, van der Most and Neumann, \cite{frysztackiInterannualElectricityDemand2024} used hourly profiles from the Open Power Systems Database \cite{muehlenpfordtTimeSeries2020} to disaggregate the daily electricity demand to an hourly resolution, on a country level. We further take this data and scale it by a factor of two, based on scenarios presented by the European Commission \cite{europeancommissionIndepthAnalysisSupport2018}, while leaving the shape of the load curve untouched.

Weather data for the performance of variable renewable energy is generated through the xarray-based Python library atlite \cite{hofmannAtliteLightweightPython2021}, which converts climate data (in our case ERA5 weather-reanalysis from ECMWF \cite{hersbachERA5GlobalReanalysis2020}) to time series in a 30x30km grid cell. With investments in variable renewable energy at a country level, as in our case, the grid cells form an average for the full spatial extent of each country. To address the fact that the average capacity factor for solar PV, onshore and offshore wind will be reduced by poorly-performing grid cells (e.g. with low wind speeds) which in reality would not be considered for the deployment of these technologies, we apply a so-called cut-off factor. The cut-off factor excludes grid cells with an average capacity factor lower than a set threshold. For solar, onshore and offshore wind, this threshold is set to 0.09, 0.15 and 0.20 respectively. Which grid cells are available to the model determine the available land area for VRE deployment, and excluded areas are primarily based on Corine Land Cover codes\footnote{\url{https://land.copernicus.eu/content/corine-land-cover-nomenclature-guidelines/html/}}. 

Hydropower plants are modelled in a simplified and aggregated manner, where each zone of the model only have one hydropower plant with the aggregated power and storage capacity. To ensure that the hydropower data is consistent with historical levels, we normalise it based on data from the U.S. Energy Information Administration \cite{u.s.energyinformationadministrationElectricity2024}. Furthermore, we base existing hydropower capacities on the JRC Hydro-power database \cite{feliceJRCHydropowerDatabase2021}.

Although previous studies have shown the issue of weather year variability (see e.g. Refs. \cite{grochowiczIntersectingNearoptimalSpaces2023,pfenningerLongtermPatternsEuropean2016,staffellIncreasingImpactWeather2018}) and that using a single year of weather data may largely skew the results, we only utilise historical data from 2010 as the main focus of this work is on differences in analysis given different interpretations of distributive justice.

The model does not have access to any negative emission technologies, such as bioenergy carbon capture and storage (BECCS) or direct air capture (DAC), and is constrained to emit on average 2gCO\textsubscript{2}/kWh generated, which is meant to represent the contribution of the electricity system in a net-zero energy system \cite{committeeSixthCarbonBudget2020}. This constraint is system-wide and does not include any detail on the contribution of individual zones within the model, which may lead to uneven mitigation patterns across the system. 

The baseline for cross-border transmission capacities is based on reported historical interconnection from ENTSO-E as well planned new interconnectors from figure 3.1 in the Ten-Year Network Development Plan 2020 \cite{entso-eTenYearNetworkDevelopment2021}. To allow some flexibility towards 2050, we allow for a three-fold increase in capacities. 

\begin{table*}[h!]
\centering
\label{tab:tech_costs}
\caption{Technology cost parameters in 2024€}
\begin{tabular*}{\tblwidth}{@{}LCCC@{}}
\toprule
Technology & Overnight CAPEX [€k/MW] & FOM [€k/MW] & VOM [€k/MWh] \\
\midrule
Solar PV         & 398.1   & 8.19    & 0.0016 \\
Onshore wind     & 1220.5   & 32.8   & 0.0082 \\
Offshore wind    & 2193.6  & 139.3   & 0.0033 \\
Nuclear          & 6507.1  & 103.4  & 0.0066 \\
HydroRoR         & -  & 68.3 & 0.0033 \\
HydroRES         & -  & 68.3 & 0.0033 \\
Gas with CCS     & 1931.4  & 54.1 & 0.0016 \\
Gas w/o CCS      & 506.2   & 22.9 & 0.0016 \\
\bottomrule
\end{tabular*}
\end{table*}

The technology cost parameters, shown in table \ref{tab:tech_costs}, are uniform across the system, meaning that there is no differentiation between the spatial zones in the model. They are based on \cite{priceRoleNewNuclear2023}, but has been converted to 2024 EUR instead of GBP. 

The complete model formulation, additional techno-economic assumptions, data and necessary code to replicate both modelling and analysis can be found in the associated GitHub and Zenodo repository\footnote{https://github.com/celeryroastalpenglow/MENOFS}. 

\section{Descriptive statistics of input data and model results}
\label{app:desc_stat}
This section contains some background and general descriptions of the input data and model results. 
\begin{figure}
    \centering
    \includegraphics[width=\textwidth]{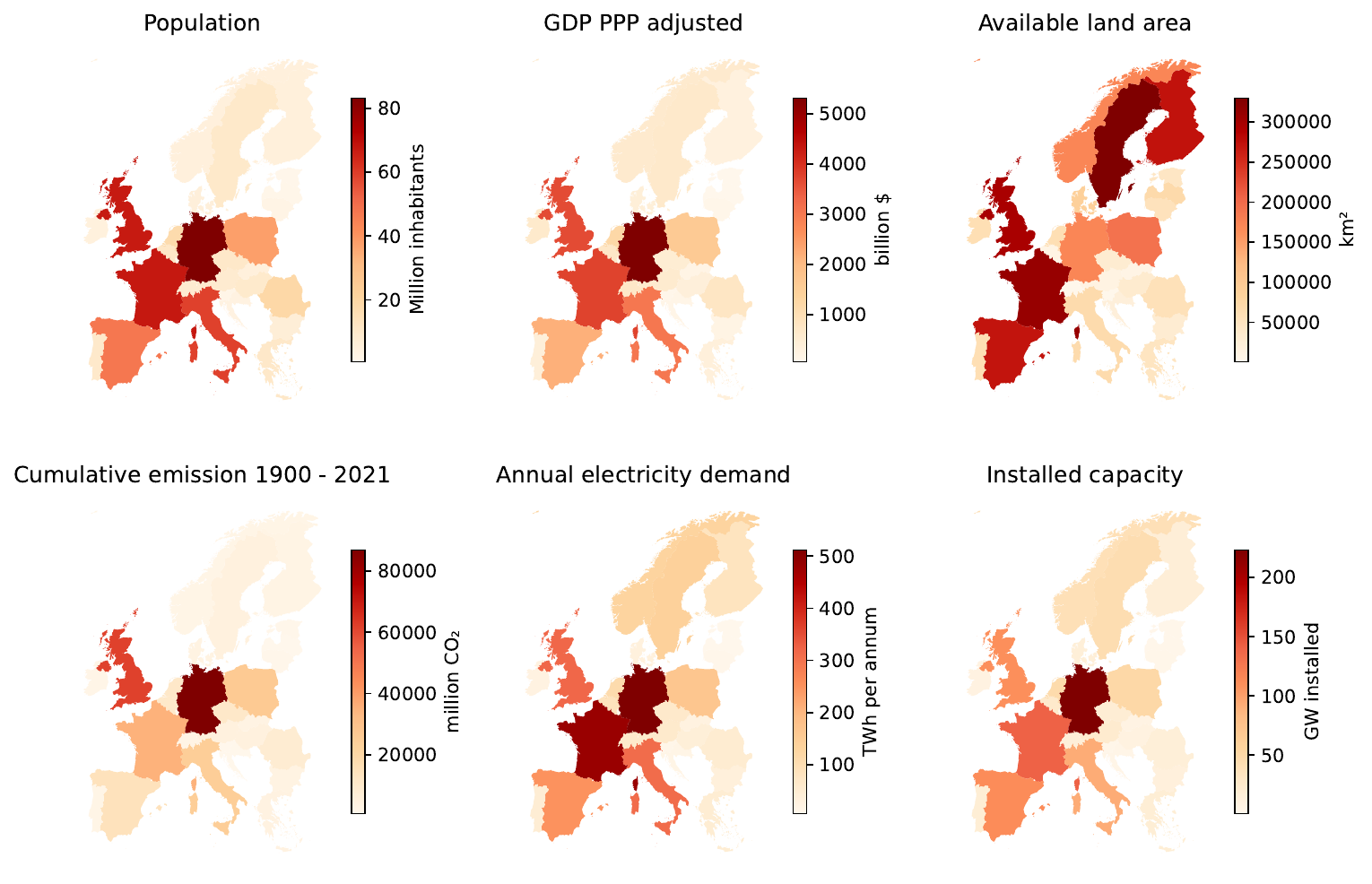}
    \caption{Spatial distribution of relevant denominators}
    \label{fig:heat_stats}
\end{figure}

In figure \ref{fig:heat_stats}, we see the spatial distribution of the differently defined denominators, which the equity factors should be distributed based on, across the spatial regions in the model. This provides some initial insight into how the factors vary across countries, and who should take a larger share of the responsibility, according to that specific denominator. 

\begin{table*}[width=.9\textwidth,cols=4,pos=h]
\label{tbl:desc_stat}
  \caption{Descriptive statistics of the Gini coefficients of each definition of justice}
  \begin{tabular*}{\tblwidth}{@{} LLLLLLLL@{}}
   \toprule
        & Max & Min & Range & Mean & Median & $\sigma$ & IQR \\
   \midrule
    Self-sufficiency    & 0.140 & 0.066 & 0.075 & 0.087 & 0.083 & 0.013 & 0.015 \\
    Capability          & 0.512 & 0.168 & 0.366 & 0.225 & 0.217 & 0.037 & 0.012 \\
    Equality            & 0.517 & 0.165 & 0.352 & 0.226 & 0.218 & 0.036 & 0.011 \\
    Land-burden         & 0.696 & 0.375 & 0.321 & 0.504 & 0.502 & 0.044 & 0.048 \\
    Responsibility      & 0.618 & 0.294 & 0.349 & 0.384 & 0.377 & 0.030 & 0.006 \\
    Grandfathering      & 0.507 & 0.169 & 0.339 & 0.250 & 0.242 & 0.034 & 0.007 \\
   \bottomrule
  \end{tabular*}
\end{table*}

Table \ref{tbl:desc_stat} provides some supplementary information to figure \ref{fig:boxplot}, with details of the associated descriptive statistics. 

\begin{figure}
    \centering
    \includegraphics[width=0.9\linewidth]{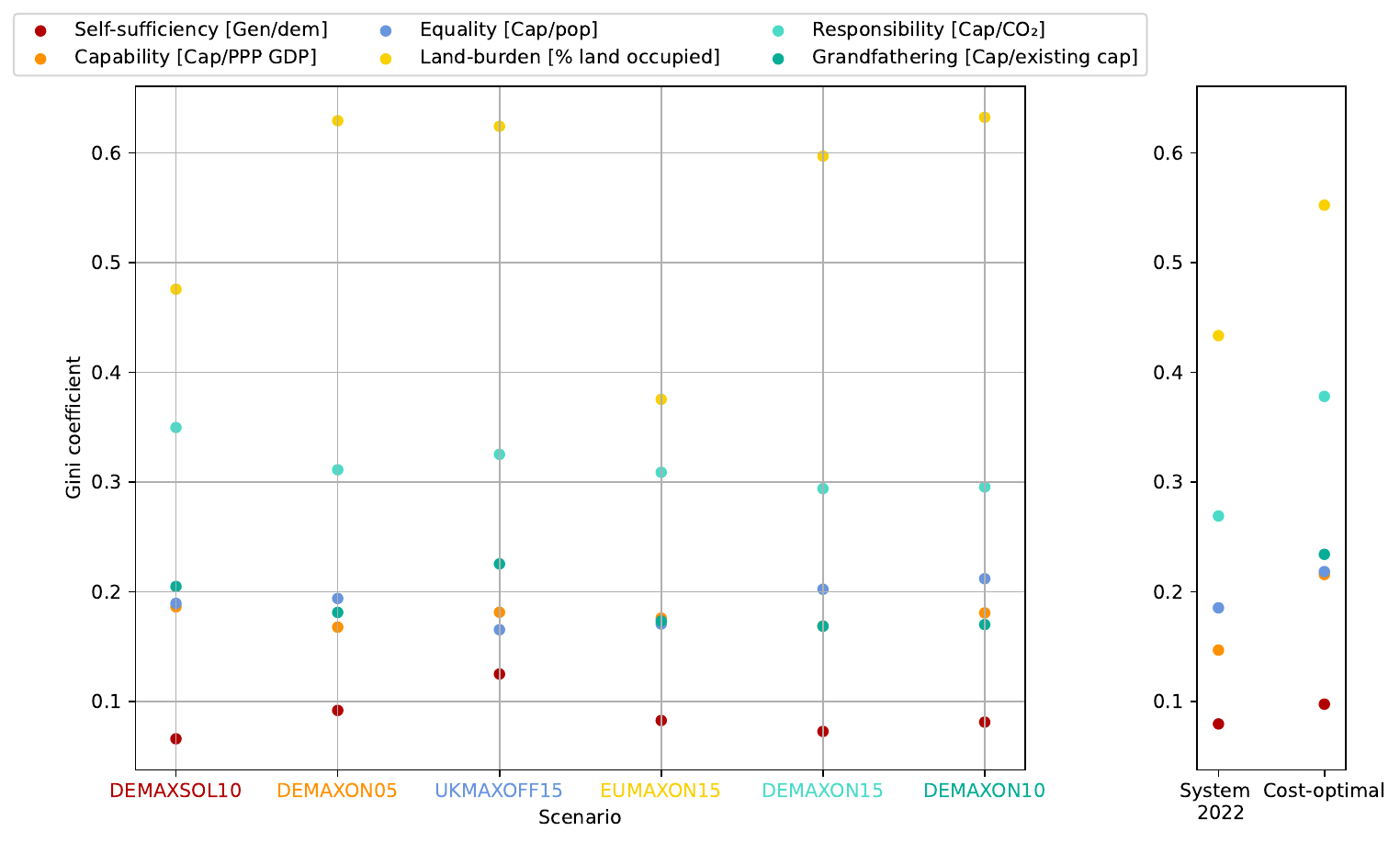}
    \caption{Gini coefficients of the top performing MGA scenarios, for each justice principle. The colour of the x-axis label illustrates which principle each scenario is a top performer for. Values close to 0 indicate high equity, whereas values close to 1 indicate high inequity.}
    \label{fig:top_perf}
\end{figure}

Figure \ref{fig:top_perf} shows the MGA scenarios which perform best (has the lowest Gini coefficient) for each of the different justice principles. It also includes a comparison with the different Gini coefficients for the 2022 system and the cost-optimal model run. 

\printcredits

\section*{Acknowledgement}
The results of this project are derived from Include – Research centre for socially inclusive energy transitions, funded by the Research Council of Norway, project no 294687. The funding institution had no involvement in any part of the study's design, and the responsibility of the work lies with the authors.

\bibliographystyle{elsarticle-num-names}

\bibliography{MENOFS}

\end{document}